\documentclass[12pt]{iopart}

\usepackage{graphicx}
\usepackage{epsfig}
\usepackage{verbatim}
\usepackage{amssymb}
\usepackage{bm}
\usepackage[T1]{fontenc}
\usepackage{eufrak}

\newlength{\bilderlength}

\newlength{\figsize}
\newcommand{\ud}{\mathrm{d}}
\newcommand{\espe}[2]{\mathbb{E}_{#1} \left[ #2 \right]}

\begin{document}

\title{Surface Tension in Kac Glass Models}
\author{Elia Zarinelli and Silvio Franz}
\address{Laboratoire de Physique Th\'eorique et Mod\`eles Statistiques\\
Universit\'e Paris-Sud 11, Centre scientifique d'Orsay,\\
15 rue Georges Cl\'emenceau 91405 Orsay cedex, France}
\ead{elia.zarinelli@lptms.u-psud.fr, \\silvio.franz@lptms.u-psud.fr}

\begin{abstract} In this paper we study a distance-dependent surface tension, defined as the free-energy cost to put metastable states at a given distance. This will be done in the framework of a disordered microscopic model with Kac interactions that can be solved in the mean-field limit.
\end{abstract}

\maketitle

\section{Introduction} \label{sec:intro}

The introduction of \emph{point--to--set} correlation functions \cite{bib:biroli, bib:montsem, bib:cavagna1, bib:cavagna2, bib:frmo} allowed important progresses in understanding the growth of static correlations in supercooled liquids near the glass transition. These non-standard correlation functions measure how deeply the effect of amorphous boundary conditions penetrates within a system. In order to introduce them, let us consider a large ensemble of interacting particles becoming glassy at low temperature.  We assume that the liquid is trapped in a metastable state. We freeze the motion of all the particles outside a sphere of radius $R$. Then we let the particles inside the sphere free to move and eventually to rearrange in a different metastable state. The effect of the external particles is to create a pinning field favouring internal configurations which best match the frozen exterior. For small radius $R$, the effect of the pinning field on the interior of the sphere is strong.  In that case the sphere remains in the same state. On the contrary, for large radius $R$, the effect becomes weak and the sphere can be found in a different state. Roughly speaking, a \emph{point--to--set} correlation function measures the overlap between the initial state and the one reached after the rearrangement of the system. It has been found in numerical experiments that on lowering the temperature the effect of the amorphous boundary conditions propagates deeper into the region \cite{bib:cavagna1,bib:cavagna2}.
\\

Standard Random First Order Transition (RFOT) \cite{bib:kirk,bib:biroli} assumes that the competition between an entropy-rich state with high energy and an entropy-poor state with low energy, can explain the transition from high-overlap to low-overlap metastable states of the previous system, as the radius of the sphere is increased. As we are going to show, such a mechanism has to be reconsidered. In order to do this, let us consider, for simplicity, a Ising-like model described by an Hamiltonian $H$. We freeze a configuration $S^{\alpha}$ in a region $A$ of the system. We study the thermodynamic considering only configurations $S$ constrained to be close to $S^{\alpha}$ is $A$:  
\begin{equation}
 Z [S^{\alpha}] = \sum_S e^{-\beta H[S]} \chi_{A}[S,S^{\alpha}] \ ,
\end{equation}
where
\begin{equation}
\chi_{A}[S^1,S^2] = \left\{
\begin{array}{lll}
1 & \ \ \rm{if } \ \ S^1_i=S^2_i& \ \ \forall i \in A\\
0 & \ \ \rm{otherwise}
\end{array}  \right.   .
\end{equation}
The thermodynamic average of an observable $\mathcal{O}$ of the system is obtained by averaging with constrained Boltzmann measure the configurations inside the sphere and with Boltzmann measure the configurations $S^{\alpha}$:
\begin{equation}
\langle \mathcal{O} \rangle =  \sum_{S^{\alpha}} \frac{e^{-\beta H[S^{\alpha}]}}{Z} \sum_{S} \chi_{A}[S,S^{\alpha}]\frac{e^{-\beta H[S]}}{Z[S^{\alpha}]} \mathcal{O}(S) \ .
\end{equation}
This average coincides with the usual thermodynamical one: $ \frac{1}{Z} \sum_{S} e^{-\beta H[S]} \mathcal{O}(S)$. This simple fact has deep implications:  in the case in which $A$ is a sphere of radius $R$, on average, the energy per degree of freedom is independent of $R$. If, for typical choices of the position of the sphere, one finds that  two thermodynamic states coexist for a well defined value of $R$,  they will have the same energy. Possible mechanisms for coexistence should therefore have a purely entropic origin \cite{bib:FranzSemerjian}.
\\

In recent numerical experiments \cite{bib:cavagna3} the energy paid to put different metastable states in contact has been measured. The procedure is the following: freeze two states $\alpha$ and $\beta$, exchange a sphere of the state $\alpha$ with a sphere of the state $\beta$ and let the system evolve. Inspired by this idea, in the present work we want to introduce a different \emph{point--to--set} correlation function defined as the free-energy cost to put different metastable states at distance $l$. In order to do that, we consider a \emph{sandwich}-geometry: two regions of the space divided by a box of width $l$ and then freeze different metastable states at opposite sides of the box, Figure \ref{fig:sandwich}. This system is well suited in order to be studied in the framework of a $p$-spin model with Kac interaction \cite{bib:frmo,bib:frton1,bib:frton2,bib:frton3,bib:fr1,bib:fr2}.
\\

The paper is organized as follows: in Section \ref{sec:model} we introduce the model that we consider and the basic definitions; in Section \ref{sec:calc} we briefly illustrate how to obtain the free energy of the system; more details on these calculations can be found in \ref{appa} and in \ref{appb}; in Section \ref{sec:res} we present our results and in Section \ref{sec:conc} we draw our conclusions.

\section{The model} \label{sec:model}

We consider a finite-dimensional version of the spherical $p$-spin model, defined on a $d$-dimensional cubic lattice $\Lambda$ of linear size $L$, whose elementary degrees of freedom are spins $S_i \in \mathbb{R}$  with $i\in \Lambda$. We introduce the interaction range $\gamma^{-1}>0$ and a non negative rapidly decreasing function $\psi(x)$ normalized by: $\int \ud^d x \psi(|x|)=1$. We define the local overlap of two configurations $S^1$ and $S^2$ as:
\begin{equation}
Q_{S^1S^2}(i)=\gamma^d \sum_{j\in \Lambda}\psi(\gamma |i-j|)S^1_j S^2_j \ .
\end{equation} 
We impose that configurations are subjected to the local spherical constraint: $Q_{S^1S^1}(i)=1$ $\forall i \in \Lambda$. We then introduce the finite-range $p$-spin Hamiltonian:
\begin{equation} 
H_p [S,J]= -\sum_{i_1,...,i_p} J_{i_1,...,i_p}S_{i_1}...S_{i_p}
\end{equation}
where the couplings $J_{i_1,...,i_p}$ are i.i.d. random variables with zero mean and variance:
\begin{equation}
\espe{}{J_{i_1,...,i_p}^2}=\gamma^{pd}\sum_{k\in \Lambda} \psi(\gamma |i_1-k|)...\psi(\gamma |i_p-k|) \ .
\end{equation}
$\gamma^{-1}$ is the interaction range since only variables located at vertices $i$ and $j$ such that $|i-j|<\gamma ^{-1}$ really interact. This also implies that the Hamiltonian is a random variable with zero mean and variance:
\begin{equation}
\espe{}{H[S^1,J]H[S^2,J]}=\sum_{i\in \Lambda}f(Q_{S^1S^2}(i)) \ ,
\end{equation}
where $f(x)$ is a polynomial with positive coefficients, for example $f(x)=x^p$, if we consider a pure $p$-spin model; in the following we will consider $f(x)=\frac{1}{10}x^2+x^4$, where the quartic term assures to have a regular gradient expansion of the free-energy density. 

We analyze the model in the Kac limit: $L, \gamma^{-1} \to \infty$ with $L \gg \gamma^{-1}$, where the model can be solved by saddle-point approximation.
\\

\begin{center}
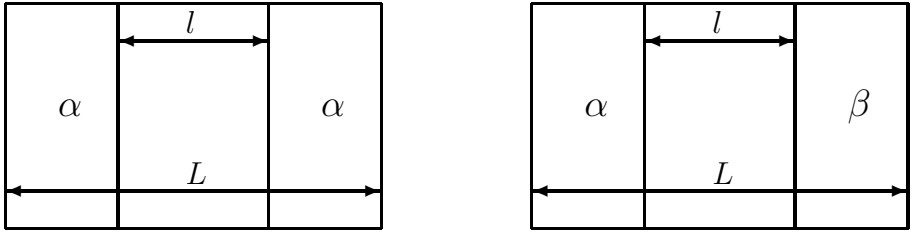
\begin{figure}[htbp]
\setlength{\unitlength}{1cm}
\begin{picture}(14,3)
\thicklines
\put(1,0){\line(1,0){5}}
\put(6,0){\line(0,1){3}}
\put(1,3){\line(1,0){5}}
\put(1,0){\line(0,1){3}}
\put(2.5,0){\line(0,1){3}}
\put(4.5,0){\line(0,1){3}}
\put(3.5,2.5){\vector(1,0){1}}
\put(3.5,2.5){\vector(-1,0){1}}
\put(3.5,0.5){\vector(1,0){2.5}}
\put(3.5,0.5){\vector(-1,0){2.5}}
\put(3.4,0.6){$L$}
\put(3.4,2.6){$l$}
\put(1.7,1.5){\large{$\alpha$}}
\put(5.2,1.5){\large{$\alpha$}}

\put(8,0){\line(1,0){5}}
\put(13,0){\line(0,1){3}}
\put(8,3){\line(1,0){5}}
\put(8,0){\line(0,1){3}}
\put(9.5,0){\line(0,1){3}}
\put(11.5,0){\line(0,1){3}}
\put(10.5,2.5){\vector(1,0){1}}
\put(10.5,2.5){\vector(-1,0){1}}
\put(10.5,0.5){\vector(1,0){2.5}}
\put(10.5,0.5){\vector(-1,0){2.5}}
\put(10.4,0.6){$L$}
\put(10.4,2.6){$l$}
\put(8.7,1.5){\large{$\alpha$}}
\put(12.2,1.5){{\large$\beta$}}

\end{picture}
\caption{The \emph{sandwich}-geometry for a system $\alpha \alpha$ (left) and $\alpha \beta$ (right). The box $B(l)$ is the central region, $A^+(l)$ and $A^-(l)$ are the lateral ones.}
\label{fig:sandwich}
\end{figure}
\end{center}

The sandwich-geometry is implemented by considering three regions of the lattice $\Lambda$: $A^+(l)$, $A^-(l)$ and a box $B(l)$, Figure \ref{fig:sandwich}. In order to put the same or different states at opposites sides of the box, we introduce two different systems, that we call $\alpha \alpha$ and $\alpha \beta$:

\begin{itemize}
\item system $\alpha \alpha$: we fix a configuration $S^{\alpha}$ drawn from the Boltzmann equilibrium measure. We consider the thermodynamic of configurations $S$ constrained to be close to $S^{\alpha}$ both in $A^+(l)$ and in $A^-(l)$;
\item system $\alpha \beta$: we fix two configurations $S^{\alpha}$ and $S^{\beta}$ drawn from the Boltzmann equilibrium measure. We consider the thermodynamic of configurations $S$ constrained to be close to $S^{\alpha}$ in $A^+(l)$ and to $S^{\beta}$ in $A^-(l)$.
\end{itemize}

We consider a system $\alpha \beta$. Let be $\mathcal{O}$ an observable of the system and $\bar{q}\le1$. The constrained Boltzmann measure $\langle \cdot \rangle_{\alpha \beta}(l)$ is:
\begin{eqnarray} \label{eq:valmed}
\langle \mathcal{O} \rangle_{\alpha \beta}(l) \equiv & \frac{1}{Z[S^{\alpha}_{A^+},S^{\beta}_{A^-}]}  \int \ud S \mathcal{O}(S) e^{-\beta H[S,J]} \nonumber \\ & \times \prod_{i \in A^-} \delta (Q_{S^{\alpha}S}(i)- \bar{q}) \prod_{i \in A^+} \delta (Q_{S^{\beta}S}(i)- \bar{q})  
\end{eqnarray}
where $\int$ denotes integration over configurations satisfying the local spherical constraint. The partition function is:
\begin{eqnarray}
Z[S^{\alpha}_{A^+},S^{\beta}_{A^-}] \equiv & \int \ud S e^{-\beta H[S,J]} \nonumber \\ & \times \prod_{i \in A^-} \delta (Q_{S^{\alpha}S}(i)- \bar{q}) \prod_{i \in A^+} \delta (Q_{S^{\beta}S}(i)- \bar{q}) \ .
\end{eqnarray}
The symbol $\mathbb{E}$ represents the average over both the distribution of fixed configurations  $S^{\alpha}$ and $S^{\beta}$ and the disorder; the free energy of the system $F_{\alpha \beta}(l)$ is then:
\begin{equation}
F_{\alpha \beta}(l,T) \equiv -\frac{1}{\beta} \ \espe{}{ \ln Z[S^{\alpha}_{A^+},S^{\beta}_{A^-}] } \ .
\end{equation}
For a system $\alpha \alpha$, the constrained Boltzmann measure $\langle \cdot \rangle_{\alpha \alpha}(l)$ is obtained by imposing the constraint $\prod_{i \in A^+ \cup A^-} \delta (Q_{S^{\alpha}S}(i)- \bar{q})$; then $F_{\alpha \alpha}(l,T)\equiv -\frac{1}{\beta}\espe{}{ \ln Z[S^{\alpha}_{A^+ \cup A^-}] } $
\\
 
As we will see in the following, $F_{\alpha \beta}(l,T) $ and $F_{\alpha \alpha}(l,T)$ can be calculated in the Kac limit, $\gamma \to 0$ taken after $L \to \infty$.  This allows us to measure the free-energy cost per unit area to put different metastable states at a distance $l$: 
\begin{equation} \label{eq:sigma}
Y(l,T) \equiv \lim_{\gamma \to 0} \lim_{L\to \infty }\frac{F_{\alpha \beta}(l,T)-F_{\alpha \alpha}(l,T)}{L^{d-1}} \ ;
\end{equation}
this quantity can be interpreted as an effective, distance-dependent, surface tension.

\section{Calculations} \label{sec:calc}

In the following we consider a system $\alpha \beta$; a system $\alpha \alpha$ can be treated in the same way. In order to calculate $F_{\alpha \beta}$, the average $\mathbb{E}$ can be taken by introducing replicas along the lines of \cite{bib:fr1, bib:fr2} (more details on calculations can be found in \ref{appb}). Integrals over spin variables are then treated for an $(m+n) \times (m+n)$ matrix order parameter $q_{ab}(i)$. We rescale the position to define $x=i \gamma \in [-\hat{L}, \hat{L}]^d$, $\hat{L} \equiv \gamma L$ to get:
\begin{equation}
F_{\alpha \beta}(\hat{l}) = -\frac{1}{\beta} \ \lim_{m,n\to 0 } \int [\ud q_{ab}] e^{-\frac{1}{\gamma^d} \mathcal{S}_{\alpha \beta}(q_{ab})}  \ .
\end{equation} 
The dependency upon $\gamma$ is now completely explicit and, for $\gamma \to 0$,  the functional integral can be performed using the saddle-point method. We look for a replica symmetric saddle point $q^{\mathrm{RS}}_{ab}(x)$. This is characterized by three scalar functions $p_1(x)$, $p_2(x)$ and $q(x)$; $p_1$ and $p_2$ are the local overlap between the constrained configuration and the reference configuration $S^{\alpha}$ and $S^{\beta}$ respectively and $q$ is the local overlap of two constrained configurations when they belong to the same metastable state (see \ref{appa} for more details). Using this ansatz we obtain that $\mathcal{S}_{\alpha \beta}(q_{ab}) = n \int \mathcal{L}_{\alpha \beta} \ud^d x + O(n^2)$, where:
\begin{eqnarray} \label{eqlag}
\mathcal{L}_{\alpha \beta}(x)  = &  -\frac{\beta^2}{2} [  f(1)+ 2f((\psi \ast p_1)(x)) +2f((\psi \ast p_2)(x)) -f((\psi \ast q)(x))  ] + \nonumber \\ &  + \frac{1}{2} \left[  \log(1-q(x)) - \frac{p_1^2(x)+p_2^2(x)-q(x)}{1-q(x)}  \right] 
\end{eqnarray}
with:
\begin{equation}
(\psi \ast q)(x) = \int \ud^d y \psi(|y-x|)q(y) \ .
\end{equation}
The constraint enforcing $S$ to be close to $S^{\alpha}$ in $A^-(\hat{l})$ and to $S^{\beta}$ in $A^+(\hat{l})$ is fulfilled by setting $p_1(x)=\bar{q}$ for $x\in A^-(\hat{l})$ and $p_2(x)=\bar{q}$ for $x \in A^+(\hat{l})$. We obtain $F_{\alpha \beta}(\hat{l})$ by evaluating the fields $p_1(x)$, $p_2(x)$ and $q(x)$ in the saddle point of the action $\mathcal{S}^0_{\alpha \beta}=\int \ud ^d x \mathcal{L}_{\alpha \beta}(x)$. The resulting free energy will present an extensive part $O(L^d)$ which will be the same for a system $\alpha \alpha$ and for a system $\alpha \beta$.  Then, in the calculation of the surface tension $Y(\hat{l},T)$, the extensive part of the free energy will erase and contributions come only from the sub-leading order $O(L^{d-1})$; the resulting form of the surface tension is $Y(\hat{l},T) = \hat{F}_{\alpha \beta}(\hat{l}, T)- \hat{F}_{\alpha \alpha}(\hat{l}, T)$, where $\hat{F}_{\alpha \beta}(\hat{l}, T) = \frac{1}{\beta}\int_0^{\hat{l}} \ud x\mathcal{L}_{\alpha \beta}(x)$.
\\

We introduce a simplification in the Lagrangians: we expand the terms of the form $f((\psi \ast q)(x))$ in gradient of $q(x)$ and we truncate to the second order obtaining $f(q(x))-cf''(q(x)) (\nabla q)^2(x)$ where $c=\frac{1}{2d}\int z^2 \psi(|z|) \ud z^d$ (in our running example $c=1$). We find the saddle-point fields iterating numerically the Euler-Lagrange equations of (\ref{eqlag}).

\section{Results} \label{sec:res}

The system $\alpha \alpha$ has been studied in spherical geometry \cite{bib:frmo}; we verified that in sandwich-geometry the behaviour does not change with respect to the spherical one. Two critical temperatures characterize the system: $T_s \approx 0.766287$ and $T_d \approx 0.813526$. 

Setting the temperature of the system $T  \gtrsim T_d$, we find two lengths: $\hat{l}_0(T)$ and $\hat{\xi}_d(T)$, such that, for widths of the box $\hat{l} \in [\hat{l}_0(T), \hat{\xi}_d(T)]$,  the action $\mathcal{S}^0_{\alpha \alpha}$ has two local minima. A minimum is characterized by a saddle-point field $p(x)$ rapidly decaying to zero in the interior of the box; we name this low-overlap minimum. The other minimum is characterized by a saddle-point field $p(x)$ everywhere large; we name this high-overlap minimum. For $\hat{l} > \hat{\xi}_d$ ($\hat{l} < \hat{l}_o$) only the low-(high-)overlap minimum exists. $\hat{\xi}_s(T)$ is defined as the minimum value of $\hat{l}$ such that the low-overlap minimum is the global minimum of the action. The critical temperatures $T_s$ and $T_d$ are defined as the temperature at which $\hat{\xi}_s(T)$ and $\hat{\xi}_d(T)$ respectively diverge. 

\begin{figure}[H]
\begin{center}
\includegraphics[width=7cm, angle=270]{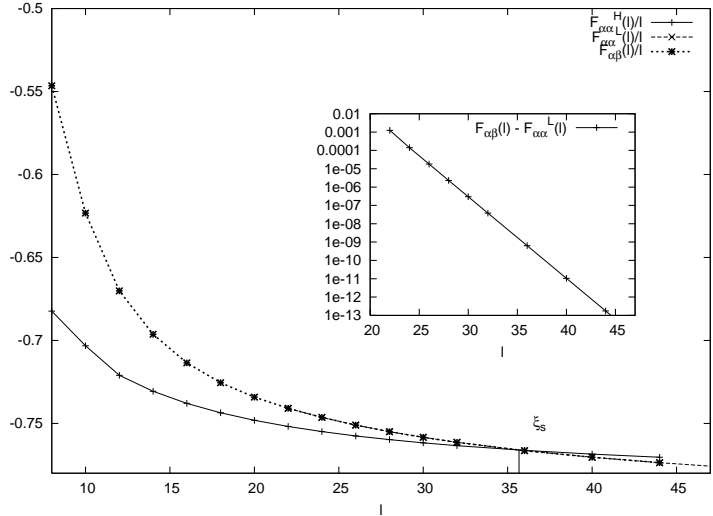}
\caption{Main figure: Plot of the sub-extensive part of the free energy divided by the size as a function of  $\hat{l}$ for a system at a temperature $T=0.7874$ of: high-overlap minimum of a system $\alpha \alpha$, $\hat{F}_{\alpha \alpha}^H(\hat{l})/ \hat{l}$; low-overlap minimum of a system $\alpha \alpha$, $\hat{F}_{\alpha \alpha}^L(\hat{l})/\hat{l}$; unique minimum of a system $\alpha \beta$, $\hat{F}_{\alpha \beta}(\hat{l})/ \hat{l}$. The static correlation length $\hat{\xi}_s$ is pointed out. Using this scale $\hat{F}_{\alpha \alpha}^L(\hat{l})$ and  $\hat{F}_{\alpha \beta}(\hat{l})$ are indistinguishable. Inset: the difference $\hat{F}_{\alpha \alpha}^L(\hat{l})- \hat{F}_{\alpha \beta}(\hat{l})$ in logarithmic scale. }
\label{fig:enlib}
\end{center}
\end{figure}

For a better comprehension, we present in Figure \ref{fig:enlib} the plot of the sub-extensive part of the free energy of high-(low-)overlap minimum $\hat{F}_{\alpha \alpha}^H(\hat{l})$ ($\hat{F}_{\alpha \alpha}^L(\hat{l})$) divided by the size $\hat{l}$ for a system at a temperature $T_s <T< T_d$. $\hat{\xi}_s(T)$ is then the value of $\hat{l}$ where $\hat{F}_{\alpha \alpha}^L(\hat{l})$ and $\hat{F}_{\alpha \alpha}^H(\hat{l})$ cross. Then the global free energy of a system $\alpha \alpha$ is  $F_{\alpha \alpha}(\hat{l})= \min \left\{  F_{\alpha \alpha}^L(\hat{l}), F_{\alpha \alpha}^H(\hat{l})\right\}$.
\\

\begin{figure}[H]
\begin{center}
\includegraphics[width=7cm, angle= -90]{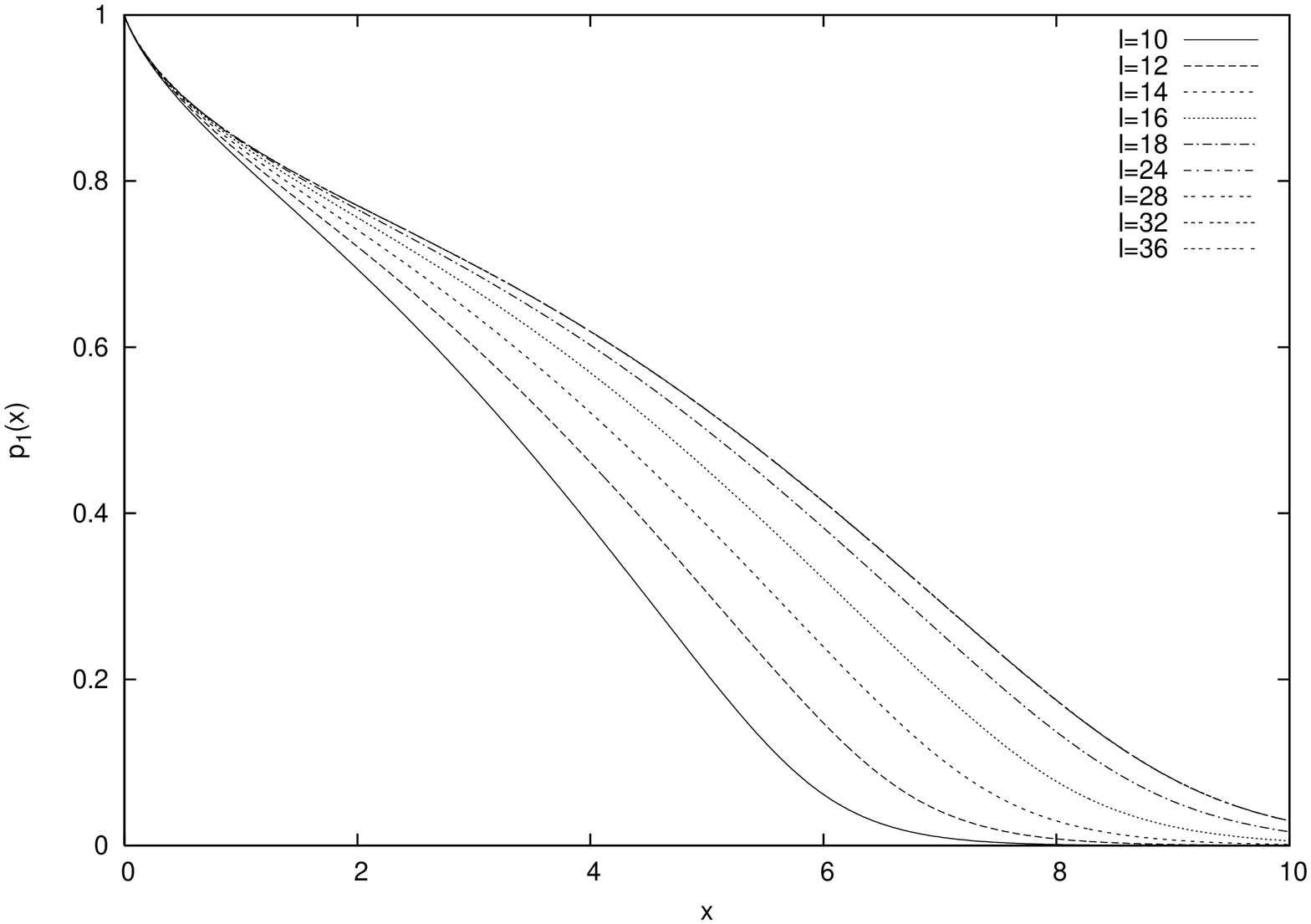}
\caption{Plot of the profiles of the saddle-point field $p_1(x)$ for a system $\alpha \beta$ at temperature $T=0.8$ for different values of the box $\hat{l}$. At this temperature $\hat{\xi}_s \sim 24$.}
\label{fig:absol}
\end{center}
\end{figure}

On the other hand, in the case of a system $\alpha \beta$, the action $\mathcal{S}^0_{\alpha \beta}$ has always a single minimum. Profiles of the saddle-point field $p_1(x)$ can be seen in Figure \ref{fig:absol}. The sub-extensive part of the free energy of the unique minimum $\hat{F}_{\alpha \beta}(\hat{l})/\hat{l}$ for a temperature $T_s< T< T_d$ is also plotted in Figure \ref{fig:enlib}. At all temperatures and values of $\hat{l}$ that we have studied, the sub-extensive part of the free energy of a system $\alpha \beta$ $\hat{F}_{\alpha \beta}(\hat{l})$ is close to the sub-extensive part of the low-overlap free energy of a system $\alpha \alpha$ $\hat{F}_{\alpha \alpha}(\hat{l})$, as can be seen in the inset of Figure \ref{fig:enlib}. 
\\

\begin{figure}[H]
\begin{center}
\includegraphics[width=8cm, angle=-90]{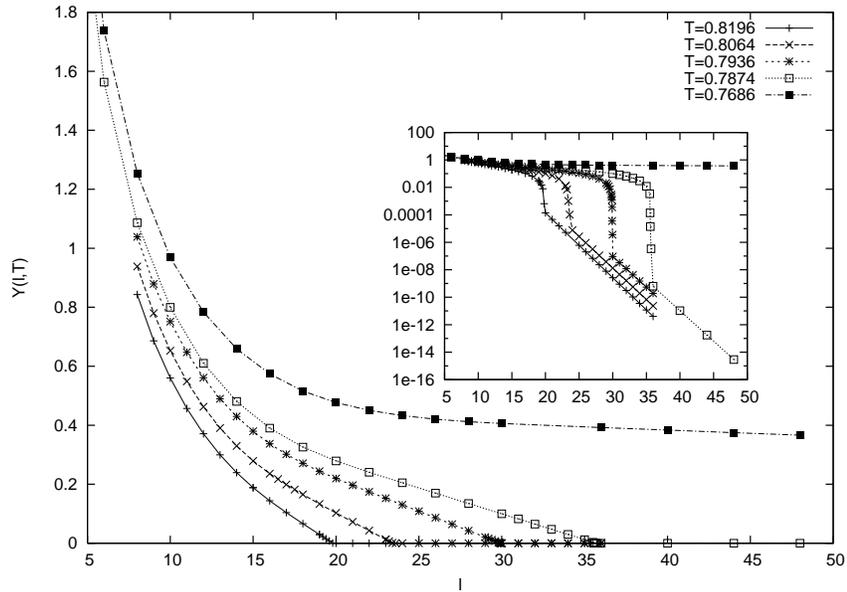}
\caption{Plot of $Y(l,T)$ for different temperatures as a function of the width of the box $\hat{l}$. We remember that $T_s \approx 0.766287$ and $T_d \approx 0.813526$.}
\label{fig:diff}
\end{center}
\end{figure}

In Figure \ref{fig:diff} we follow the evolution of $\hat{l}$-dependent surface tension $Y(\hat{l},T)$ for systems at different temperatures $T>T_s$. We note that the static correlation length $\hat{\xi}_s(T)$ separates two regimes. For $\hat{l}<\hat{\xi}_s(T)$,  $Y(\hat{l},T)$ has a power-law followed by a linear decrease. For $\hat{l}>\hat{\xi}_s(T)$, as we see in the inset of Figure \ref{fig:diff}, the decrease becomes exponential:
\begin{equation}
Y(\hat{l},T) \sim C \ e^{-\hat{l}/\tilde{l}},
\end{equation}
with $\tilde{l}$ weakly dependent on the temperature and showing no evident relation with $\hat{\xi}_s$. This shows that the surface tension $Y(\hat{l},T)$ is sensibly different from zero only for $\hat{l}\lesssim \hat{\xi}_s$. A similar result has been obtained in \cite{bib:moore}; in that case the interface free energy has been obtained changing the boundary conditions along one direction, from periodic to anti-periodic.    

Particular attention must be spent in the case $T=T_s$. At $T_s$, the static correlation length $\hat{\xi}_s$ diverges. This means that the high-overlap minimum is the global minimum of the action $\mathcal{S}^0_{\alpha \alpha}$ for all the  values of $\hat{l}$. We see in Figure \ref{fig:diff} that, for $T$ approaching $T_s$, the profile of $Y(\hat{l},T)$ takes the shape of a plateau. Consequently, at the critical temperature $T_s$, in the limit $\hat{l} \to \infty$, the surface tension $Y(\hat{l},T_s)$ does not fall to zero and takes a limiting value $Y(T_s)$. Arguably, the value $Y(T)$ is different from zero for temperatures $T<T_s$. 
\\

According to phenomenological arguments \cite{bib:biroli}, the static correlation length $\hat{\xi}_s(T)$ can be interpreted as the typical size of metastable states of a system at a temperature $T$. Following this idea, in a system $\alpha \beta$ we are freezing a patchwork of metastable states of size $\hat{\xi}_s(T)$ outside the box and letting the system free to rearrange inside the box. If the width of the box is larger than the typical metastable-state size, $\hat{l} \gg \hat{\xi}_s(T)$, the system inside the box has enough space to rearrange in many different metastable states. On the contrary, when the width of the box is smaller than the metastable-state size,  $\hat{l}<\hat{\xi}_s$, since there is not enough space to create metastable states on the interior, the frozen states are in contact and then ``repel" each other.  This explains why the surface tension $Y(\hat{l},T)$ is significantly different from zero only for $\hat{l} < \hat{\xi}_s(T)$ and why the overlap profiles $p_1(x)$ and $p_2(x)$ between frozen metastable states and the interior of the box decrease faster for small boxes. At the critical temperature $T_s$ the size of metastable states diverges. Consequently, the surface tension takes a finite value also in the limit $\hat{l} \to \infty$.  
\\

\begin{figure}[H]
\begin{center}
\includegraphics[width=8cm, angle= -90]{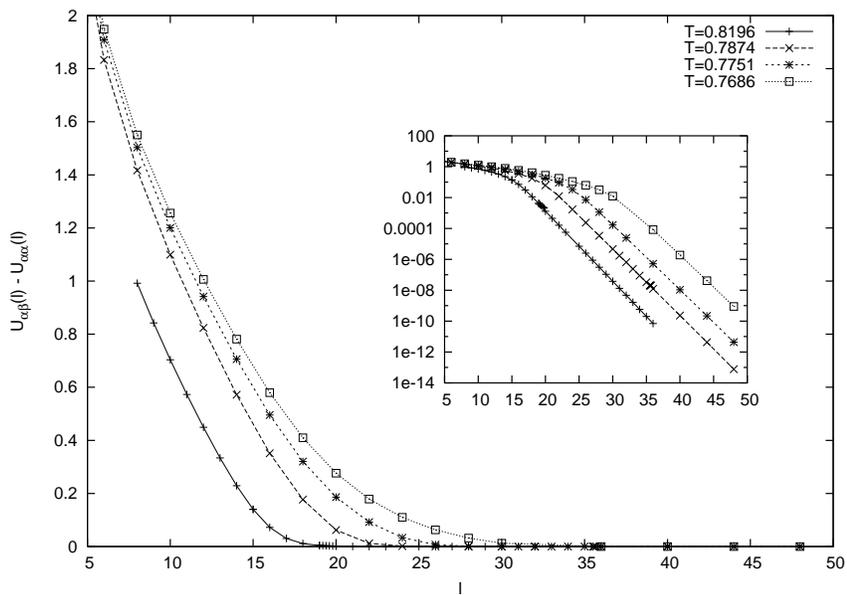}
\caption{Plot of $U_{\alpha \beta}(\hat{l})-U_{\alpha \alpha}(\hat{l})$ for different temperatures as a function of the width of the box $\hat{l}$. We remember that $T_s \approx 0.766287$ and $T_d \approx 0.813526$.} 
\label{fig:enint}
\end{center}
\end{figure}

Other observables of the system have been considered. We studied the internal energy $U$. We verified that for a system $\alpha \alpha$ the high-overlap and the low-overlap phases have the same energy, as motivated in Section \ref{sec:intro}. In Figure \ref{fig:enint} we follow the evolution of $U_{\alpha \beta}(\hat{l})-U_{\alpha \alpha}(\hat{l})$ for different temperatures of the system. A detailed derivation of this quantity can be found in \ref{appb}. In this case, we note a power-law followed by a an exponential decrease. 
\\

\begin{figure}[H]
\begin{center}
\includegraphics[width=7cm,angle=-90]{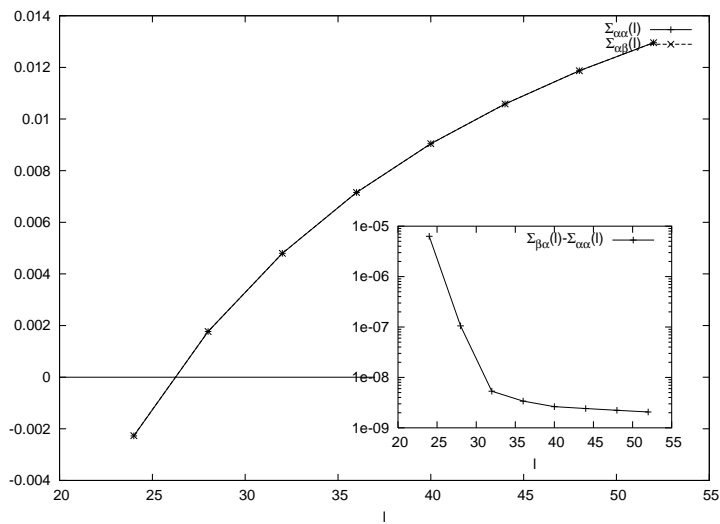}
\caption{Main figure: The configurational entropy $\Sigma_{\alpha \alpha}(l)$ in function of $l$ for a system $\alpha \alpha$ and $\Sigma_{\alpha \beta}(l)$ for a system $\alpha \beta$ at a temperature $T=0.8$. Inset: the difference $\Sigma_{\alpha \beta}(l)-\Sigma_{\alpha \alpha}(l)$.}
\label{fig:entropia}
\end{center}
\end{figure}

We also computed the configurational entropy $\Sigma$ as a function of the size $\hat{l}$ of the box, Figure \ref{fig:entropia}. For a system $\alpha \alpha$ only the low-overlap phase presents a configurational entropy $\Sigma_{\alpha \alpha}$ different from zero. As noticed in \cite{bib:frmo}, for $\hat{l}<\hat{l}^{1RSB}$ the replica-symmetric solution is incorrect since it gives a negative entropy. We found that the same is true for a system $\alpha \beta$. In the inset of Figure \ref{fig:entropia} we plot the difference between the configurational entropy of the two systems. We note that this quantity is a decreasing function of the size $\hat{l}$ of the system. This is consistent with the observation that the system loses memory of the frozen exterior for large sizes of the box.

\section{Conclusions}\label{sec:conc}

In this paper we have studied a distance-dependent surface tension, defined as the free-energy cost to put metastable states at a given distance. This has been done in the framework of a disordered microscopic model with Kac interactions that can be solved in the mean-field limit. We have found that the surface tension is sensibly different from zero only for distances between metastable states smaller than the static correlation length of the system. A description of this behaviour in terms of a phenomenological droplet argument has been proposed. Other observables, like the internal energy and the configurational entropy, has been studied. The behaviour of the configurational entropy allowed to identify under which size the replica-symmetric ansatz becomes incorrect and a 1-RSB solution must be considered.    

\ack It is a pleasure to thank D. Fichera, M. Castellana and G. Biroli for interesting discussions.

\appendix 
\section{}\label{appa}
We give an explicit formulation of the overlap matrix in the replica symmetric ansatz $q^{\mathrm{RS}}_{ab}(x)$; the overlap matrix is an $(m+n) \times (m+n)$ matrix with $m$ and $n$ real numbers; if we take $r$ and $n$ integer, we can visualize the matrix in the following way:
\begin{equation}
q^{\mathrm{RS}}=\left[ \begin{array}{cc}
A & B \\
B^T & C 
\end{array} \right] \ .
\end{equation} 
The $n \times n$ matrix $C$ is the overlap matrix between configuration that are taken with constrained Boltzmann measure and subjected to local spherical constraint; the replica symmetric ansatz imposes that $C_{ab}=q(x)$ for all $a \neq b$ and the spherical constraint that $C_{aa}=1$; then $C$ can be written in the form: 

\begin{equation}
C=\left[ \begin{array}{ccccc}
1 & q(x) & q(x)& \dots &q(x) \\
q(x) & 1&q(x) & \dots & q(x) \\
\dots & \dots & \dots & \dots \\
q(x) & q(x) &q(x)& \dots &1
\end{array} \right] \ .
\end{equation}
The $m \times m$ $A$ matrix is the overlap matrix between configuration that are taken with Boltzmann measure and subjected to local spherical constraint; we impose the out of diagonal elements equal to zero, then we obtain that $A$ is the identity matrix: $A=1$.

The $m \times n$ $B$ matrix is the overlap matrix between configuration that are taken with Boltzmann measure and configuration that are taken with constrained Boltzmann measure; we impose all the elements of this matrix equal to zero, except the last two lines that are equal to $p_1(x)$ and to $p_2(x)$; then $A$ can be written in the form:
\begin{equation}
B=\left[ \begin{array}{cccc}
0 & 0 &  \dots & 0 \\
0 & 0& \dots & 0 \\
\dots & \dots & \dots & \dots \\
p_1(x) & p_1(x) & \dots & p_1(x) \\
p_2(x) & p_2(x) & \dots & p_2(x)
\end{array} \right] \ .
\end{equation}

\section{}\label{appb}
The internal energy of a system $\alpha \beta$ is:
\begin{equation*}
U_{\alpha \beta}(l) \equiv \espe{}{\langle H[s,J] \rangle_{\alpha \beta}(l)}  =  \espe{}{-\frac{\partial}{\partial \beta} \ln Z[s^{\alpha}_{A^+},s^{\beta}_{A^-}] } \nonumber 
\end{equation*} 
We give an explicit derivation of $U_{\alpha \beta}(l)$; similar calculations allow to obtain the free energy $F_{\alpha \beta}(l)$. We introduce two different temperatures $\beta_1$ and $\beta_2$ and $n$ and $m$ replicas of the system;
\footnotesize
\begin{eqnarray}
U_{\alpha \beta}(l) & =  & \espe{}{\espe{s^{\alpha}}{ \espe{s^{\beta}}{-\frac{\partial}{\partial \beta} \ln Z[s^{\alpha}_{A^+},s^{\beta}_{A^-}] }  }}  \nonumber \\
 &   =  & -\frac{\partial}{\partial \beta_2} \espe{}{  \frac{1}{Z^2[\beta_1]} \int \ud s^{\alpha} \ud s^{\beta} \exp{ \left[ -\beta_1 \left( H[s^{\alpha},J] + H[s^{\beta},J]  \right) \right] } \ln{Z[s^{\alpha},s^{\beta},\beta_2]}    }   \nonumber \\ 
  &   =  & \lim_{m,n \to 0} \frac{1}{n} \left( -\frac{\partial}{\partial \beta_2} \right)   \nonumber \\  & & \espe{}{  Z^{m-2}[\beta_1] \int \ud s^{\alpha} \ud s^{\beta} \exp{ \left[ -\beta_1 ( H[s^{\alpha},J] + H[s^{\beta},J] ) \right]} Z^n[s^{\alpha},s^{\beta},\beta_2]    }  \nonumber 
\end{eqnarray}
\normalsize
where:
\footnotesize
\begin{equation*}
\mathcal{C}= \prod_{a=m+1}^{m+n} \left[  \prod_{i \in A^+} \delta (Q_{s^{m-1}s^a}(i)- \bar{q}) \prod_{i \in A^-} \delta (Q_{s^{m}s^a}(i)- \bar{q})  \right] \ . \nonumber
\end{equation*}
\normalsize
Then performing the expectation value over the disorder, the derivative and reimposing equal the temperatures we obtain:
\footnotesize
\begin{eqnarray}
 U_{\alpha \beta}(l) =  \lim_{m,n \to 0} \frac{- \beta}{n}  & \int \prod_{a=1}^{m+n} \ud s^{a}  \mathcal{C}    \exp{ \left[ \frac{\beta^2}{2} \sum_{i \in \Lambda} \sum_{1 \le a,b \le n}  f(Q_{ab}(i))) \right] }  \nonumber \\  
   & \times  \sum_{i \in \Lambda} \left[ \sum_{a,b\in C}   f(Q_{ab}(i)) + \frac{1}{2}   \sum_{ a,b \in B }   f(Q_{ab}(i)) + \frac{1}{2} \sum_{ a,b \in B^T }  f(Q_{ab}(i)) \right]  \ . \nonumber
 \end{eqnarray} 
\normalsize
Integrals over the spin variables are then traded for an $(m+n) \times (m+n)$ matrix order parameter $q_{ab}(i)$.
\footnotesize
\begin{eqnarray}
 U_{\alpha \beta} =  \lim_{m,n \to 0} \frac{- \beta}{n}  & \int \prod_{i \in \Lambda} \prod_{a, b =1}^{m+n} q_{ab}(i) \mathcal{C}  \exp{ \left[  \sum_{i \in \Lambda} \left( \frac{\beta^2}{2} \sum_{1 \le a,b \le n} f(q_{ab}(i)))  + \frac{1}{2} \log \det q(i) \right) \right] }  \nonumber \\  
   & \times  \sum_{i \in \Lambda} \left[ \sum_{ a,b \in C}  f(q_{ab}(i)) + \frac{1}{2}  \sum_{  a,b \in B }f(q_{ab}(i)) + \frac{1}{2} \sum_{  a,b \in B^T }  f(q_{ab}(i)) \right]  \ . \nonumber
 \end{eqnarray}
\normalsize
Performing the coarse graining:
\footnotesize
\begin{eqnarray}
 U_{\alpha \beta} =  \lim_{m,n \to 0} \frac{- \beta}{n} & \int [\ud q_{ab}]  \int \ud x \left[ \sum_{a,b \in C} f((\psi \ast q)_{ab}(x)) + \frac{1}{2} \sum_{ a,b \in B, B^T } f((\psi \ast q)_{ab}(x)) \right] \nonumber \\  &  \times
    \exp{ \left[ \frac{1}{\gamma^d} \int \ud x \left(  \frac{\beta^2}{2} \sum_{1 \le a,b \le n}  f((\psi \ast q)_{ab}(x)))  + \frac{1}{2} \log \det q(x)\right) \right] } \mathcal{C} \ . \nonumber
 \end{eqnarray}
\normalsize
Using the replica symmetric matrix presented in \ref{appa} we obtain:
\footnotesize
\begin{equation*}
 U_{\alpha \beta} =  \lim_{n \to 0} \frac{- \beta}{n}  \int [\ud q] [\ud p_1] [\ud p_2]   \left[ n  \int \ud x \mathcal{H}(x) +o(n^2)\right]   \exp{ \left[  -\frac{n}{\gamma^d} \int \ud x \mathcal{L}(x) + o(n^2)\right]}  \nonumber
 \end{equation*} 
\normalsize 
where:
\footnotesize
\begin{equation*}
\mathcal{H}(x)=1+f((\psi \ast p_1)(x)) + f((\psi \ast p_2)(x)) - f((\psi \ast q)(x))  \ ; \nonumber 
\end{equation*}
\begin{eqnarray}
\mathcal{L}(x)  = &  \frac{\beta^2}{2}\left[  1+ 2f((\psi \ast p_1)(x)) +2f((\psi \ast p_2)(x))-f((\psi \ast q)(x))  \right]  + \nonumber \\ & +\frac{1}{2} \left[  \log(1-q(x)) - \frac{p_1^2(x)+p_2^2(x)-q(x)}{1-q(x)}  \right] \ . \nonumber
\end{eqnarray}
\normalsize
We evaluate the action $\mathcal{S}^0_{\alpha \beta}=\int \ud ^d x \mathcal{L}_{\alpha \beta}(x)$ in the saddle point fields $p_1$, $p_2$ and $q$ and we obtain that:
\begin{equation}
U_{\alpha \beta}(\hat{l}) = -\beta \int \ud x \mathcal{H}(x)\ .
\end{equation}

\end{document}